\documentclass[a4paper]{article}
\usepackage{graphicx}

\newcommand{\x}{\mbox{\boldmath $\xi$}}
\def\Argmin{\mathop{\rm Argmin}}
\begin{document}

\bibliographystyle{unsrt}

\title{Bayes-optimal performance in a discrete space}

\author{M. Copelli$^{1}$\footnote{email address: {\tt
mauro.copelli@luc.ac.be}}, C. Van den Broeck$^{1}$\footnote{email
address: {\tt christian.vandenbroeck@luc.ac.be}} and M.
Opper$^{2}$\footnote{email address: {\tt opperm@aston.ac.uk}} \\ \ \\
\small $^{1}$ Limburgs Universitair Centrum, B-3590 Diepenbeek, Belgium \\
\small $^{2}$ Neural Computing Research Group, Aston University, 
Birmingham B4 7ET, UK}

\maketitle

\begin{flushleft}
mechanics of model systems 
statistics
\end{flushleft}

\begin{abstract}
We study a simple model of unsupervised learning where the single 
symmetry breaking vector has binary components $\pm 1$.  We calculate 
exactly the Bayes-optimal performance of an estimator which is 
required to lie in the same discrete space.  We also show that, except 
for very special cases, such an estimator cannot be obtained by 
minimization of a class of variationally optimal potentials.
\end{abstract}

\begin{sloppypar}
Statistical mechanics techniques have been used with success to study
and understand key properties of inferential
learning~\cite{Watkin93a,Opper96b}.  This approach provides explicit
and detailed results that are in many ways complementary to the more
general results obtained by statistics.  The case of non-smooth
problems, in which the parameters that have to be estimated take
discrete values, is of particular interest.  On the one hand, many of
the results from statistics can no longer be applied, while on the
other hand, the estimation of these parameters is often a
computationally hard problem.  In this paper, we present a detailed
analysis of a simple model of unsupervised
learning~\cite{Biehl93a,Biehl94b,Watkin94a,Reimann96a,Reimann96b,VandenBroeck96},
involving a single symmetry breaking vector with binary components
$\pm 1$ and highlight the differences with the case of smooth
components.  In particular we compare the results from Gibbs learning
and Bayes learning with the ones for the best binary vector and a
vector which minimizes a variationally optimal potential.

The problem is as follows: $p$ $N$-dimensional real patterns
$\{\x^{\mu}$, $\mu=1,...,p\}$, are sampled independently from a
distribution $P(\x^{\mu}|{\bf B}) \sim \delta(\x^{\mu}\cdot\x^{\mu} -
N)\, \exp\left[ -U\left( {\bf B}\cdot\x^{\mu}/\sqrt{N} \right)
\right]$ with a single symmetry breaking direction ${\bf B}$.  The
function $U$ modulates the distribution of the patterns along ${\bf
B}$.  We will focus on the properties in the thermodynamic limit
$N\to\infty$, $p\to\infty$ with $\alpha=p/ N$ finite.  One then finds
that the normalized projection $t\equiv {\bf B}\cdot\x/\sqrt{N}$ is
distributed according to (${\cal N}$ being a normalization constant)

\begin{equation}
\label{pt1}
P^{*}(t) = \frac{\cal N}{\sqrt{2\pi}} \exp\left\{ - \frac{t^{2}}{2} - 
U(t) \right\}\; ,
\end{equation}
while projections on any direction orthogonal to ${\bf B}$ are normal. 
The case of a so-called spherical prior, in which ${\bf B}$ is chosen
at random on the sphere with radius $\sqrt{N}$, was discussed in
~\cite{Reimann96a,Reimann96b,VandenBroeck96}.  As announced earlier,
we focus here on the more complicated situation in which the
components of ${\bf B}$ take binary values $\pm 1$.  The {\em prior
distribution\/} is now given by:

\begin{equation}
\label{prior}
P({\bf B}) \equiv P_{b}({\bf B}) = 
\prod_{j=1}^{N}\left[ \frac{1}{2}\delta(B_{j}-1) + 
\frac{1}{2}\delta(B_{j}+1) \right]\; .
\end{equation}

The goal of unsupervised learning is to give an estimate ${\bf J}$ of
${\bf B}$.  One way to do so is to sample ${\bf J}$ from a Boltzmann
distribution with Hamiltonian ${\cal H}({\bf
J})=\sum_{\mu=1}^{p}V(\lambda^{\mu})$, with $\lambda^{\mu}\equiv {\bf
J}\cdot\x^{\mu}/\sqrt{N}$, at temperature $T = \beta^{-1}$, for an
appropriate choice of the ad-hoc potential $V$ \cite{Schietse95}.  The
properties of such a ${\bf J}$-vector can be extracted from the
partition function :

\begin{equation}
\label{Z}
Z=\int d{\bf J}\,  P_{b}({\bf J})\, e^{-\beta 
{\cal H}(\bf J)}\; .
\end{equation}
The latter is a fluctuating quantity due to the random choice of the
patterns, but the free energy per component $f = -(N\beta)^{-1} \ln Z$
is expected to be self-averaging in the thermodynamic limit and can
therefore be calculated by averaging over the pattern distribution
with the aid of the replica trick~\cite{Mezard87a}.  Assuming replica
symmetry (RS), one finds 

\begin{eqnarray}
\label{freeenergy}
 f  & &=   \frac{1}{\beta}
\;\mathop{\rm Extr}_{R,q,\hat{R},\hat{q}}
\Bigg\{
 \frac{1}{2}(1-q)\hat{q}
+ \hat{R}R
- \int {\cal D}z \; \ln\cosh\left(z\sqrt{\hat{q}} + \hat{R}\right)
\\
& & 
- {\alpha}
\int {\cal D}^{*}t\,\int{\cal D}t'\; 
\ln\int\frac{d\lambda}{\sqrt{2\pi(1-q)}}
\exp\bigg(
-\beta V(\lambda) 
-\frac{(\lambda - t'\sqrt{q-R^{2}}-tR)^{2}}{2(1-q)}
\bigg)\Bigg\} \nonumber\; .
\end{eqnarray}
where ${\cal D}^{*}t = dt\,P^{*}(t)$ and ${\cal D}t' = dt'\,
(2\pi)^{-1/2}\exp(-t'^{2}/2)$.  The extremum operator gives saddle
point equations which determine the self-averaging value of the order
parameters.  As usual $q$ can be interpreted as the typical mutual
overlap between two samples ${\bf J}$ and ${\bf J'}$, $q={\bf
J}\cdot{\bf J'}/N$, while the performance $R$ measures the proximity
between the estimate ${\bf J}$ and the ``true'' direction ${\bf B}$,
$R={\bf J}\cdot{\bf B}/N$.  For even functions $U$, there is no
distinction between ${\bf B}$ and $-{\bf B}$, and a symmetry $R\to -R$
arises.  In the following, only $R\geq 0$ will be considered.

As a first application of eq.~\ref{freeenergy}, we turn to Gibbs
learning~\cite{Gyorgyi90,Watkin93b,Watkin94a}.  It corresponds to
sampling from the posterior distribution and is realized by taking
$\beta = 1$ and $V = U$ in eq.~\ref{freeenergy} (for more details,
see~\cite{Reimann96a}; for the estimation of $U$,
see~\cite{Reimann97a}).  In agreement with the fact that one cannot
make a statistical distinction between ${\bf B}$ and its Gibbsian
estimate ${\bf J}$, one finds that the order parameters satisfy $q_{G}
= R_{G}$ and $\hat{q}_{G} = \hat{R}_{G}$, where the subscript ${G}$
refers to Gibbs learning.  This observation allows to simplify the
saddle point equations further, and the Gibbs overlap is found to obey
the following equation:

\begin{equation}
\label{spRG}
R_{G} = F_{B}^{2}\left({\cal F}\left(\sqrt{R_{G}}\right)\right) 
\end{equation}
with
\begin{equation} 
\label{FB}
F_{B}(x)  =  \sqrt{
\int {\cal D}z\, \tanh\left( 
zx + x^{2}
\right)}
\; \;
\mbox{and}\;\;
{\cal F}\left(R\right) = 
\sqrt{
\alpha \int{\cal D}t\, 
\frac{Y^{2}(t;R)}{X(t;R)}
}
\end{equation}
and
\begin{eqnarray}
\label{X}
X(t;R)  =  \int {\cal D}t^{\prime}\,{\cal N}e^{-U(Rt + 
\sqrt{1-R^{2}}t^{\prime})} \;\;\;\;
Y(t;R)  = \frac{1}{R}\frac{\partial}{\partial t} X(t;R)\; .
\end{eqnarray}
Note that $F_{B}$ comes from the entropic term of the free energy and 
does not depend on $U$, as opposed to ${\cal F}$, $X$ and $Y$.

For $R_{G}$ small, one obtains from eqs.~\ref{spRG}-\ref{X}, upon 
assuming a smooth behavior as a function of $\alpha$, that ($\int 
{\cal D}^{*}t\;f(t)=\left< f(t)\right>_{*}$):

\begin{eqnarray}
\label{smallalphaG1}
\left<
t\right>_{*} \neq 0 & \Rightarrow & 
R_{G} \simeq \alpha\;\left< t\right>_{*}^{2} \\ 
\label{smallalphaG2}
\left<
t\right>_{*} = 0 & \Rightarrow & 
R_{G} \left\{
\begin{array}{ll}
= 0, & \alpha\leq\alpha_{G} \\
\simeq C (\alpha-\alpha_{G}), & \alpha\geq\alpha_{G}
\end{array}
\right.
\end{eqnarray}
with critical load $\alpha_{G} = \left(1 - \left<
t^2\right>_{*}\right)^{-2}$.  These results are identical to those for
a spherical prior~\cite{VandenBroeck96}.  In particular, one observes
the appearance of {\it retarded learning} when the distribution has a
zero mean along the symmetry breaking axis.  In the regime $R_{G}
\rightarrow 1$, on the other hand, one finds an exponential approach :

\begin{equation}
\label{largealphaG}
1- R_{G}(\alpha)\stackrel{\alpha\to\infty}{\simeq} 
\sqrt{\frac{\pi}{2\alpha\left<(U^{\prime})^{2}\right>_{*}}} 
\exp \left(
\frac{ -\alpha \left<(U^{\prime})^{2}\right>_{*}}{2}\right) \; ,
\end{equation}
where $U^{\prime}\equiv dU(t)/dt$.  This is now different from the
case of a spherical prior, where the approach is following an inverse
power law $1-R_{G}\sim \alpha^{-1}$ \cite{VandenBroeck96}.  The
difference becomes even more pronounced when $U$ has singular
derivatives, as is typically the case when a supervised problem is
mapped onto an unsupervised version~\cite{Reimann96a}.  Then one finds
that $R_{G}=1$ is attained at a {\it finite\/} value of $\alpha$ while
$1-R_{G}\sim \alpha^{-2}$ for a spherical prior, see~\cite{Gyorgyi90}
and~\cite{Reimann96a} for an explicit example.

Apart from its intrinsic interest, Gibbs learning is also directly
related to the Bayes optimal overlap by $R_{B}= \sqrt{R_{G}}$,
see~\cite{Opper91a,Watkin93b,Watkin94a}.  This overlap is realized by
the center of mass ${\bf J}_{B}$ of the Gibbs ensemble.  A simple
reasoning~\cite{Watkin93b,Watkin94a} shows that ${\bf J}_{B}$
maximizes the overlap $R$ averaged over the posterior distribution of
${\bf B}$.  In order to exclude the case ${\bf J}_{B}=0$ (which would
follow in the presence of the symmetry ${\bf B}\to -{\bf B}$), we will
implicitly assume an infinitesimally small symmetry breaking field in
the Gibbs distribution.

Using the self-averaging of the mutual overlap, with $q_G=R_G$, the
explicit form of ${\bf J}_{B}$ is found to be ${\bf J}_{B} =
R_G^{-1/2} Z^{-1}\int d{\bf J}\, P_{b}({\bf J})\, {\bf J}\, \exp\{
-\sum_{\mu}U(\lambda^{\mu})\}$.  In general, the components of this
center of mass are continuous, while our prime interest here is in the
optimal performance attainable by a binary vector.  The latter vector,
which we will denote by ${\bf J}_{bb}$ (for best binary), can
fortunately be easily obtained~\cite{Watkin93a}: it is the clipped
version of the center of mass ${\bf J}_{B}$, with components $({\bf
J}_{bb})_j = \mbox{sign}\left(({\bf J}_{B})_j\right)$.

To evaluate the overlap between ${\bf J}_{bb}$ and ${\bf B}$, we
recall the following general result for the overlap
$\tilde{R}=\tilde{\bf J}\cdot {\bf B}/N$ of a vector $\tilde{\bf J}$
with transformed components $\tilde{J}_{i} =
\sqrt{N}g(J_{i})/\sqrt{\sum_{i}g^{2}(J_{i})}$ (with $g$ odd and ${\bf
B}$ binary) as a function of the overlap $R$ of $\bf{J}$ with {\bf B}
(see~\cite{Schietse95} for details):

\begin{equation}
\label{rtilde} 
\tilde{R} = \frac{\int P(x)\,g(x)\,dx}
{\left[\int P(x)\,g^{2}(x)\,dx
\right]^{1/2}}\; ,
\end{equation}
where $P(x)$ is the probability density for $x \equiv J_{1}B_{1}$,
which for the prior distribution eq.~\ref{prior} is independent of the
index due to the permutation symmetry among the axes.  If ${\bf J}$ is
sampled from a spherical distribution (with ${\bf B}$ binary), then
$P(x)$ is found to be a Gaussian~\cite{Schietse95} with mean $R$ and
variance $1-R^{2}$.

In order to obtain $P(x)$ corresponding to the center of mass ${\bf
J}_{B}$, we evaluate the quenched moments of $y = x\sqrt{R_G}$:

\begin{equation}
\label{mthmoment}
\left< y^{m}\right> = \left< \left( Z^{-1} \int d{\bf J}\,P_{b}({\bf 
J})\, e^{-\sum_{\mu}U(\lambda^{\mu})} J_{1}B_1 \right)^{m} \right>\; .
\end{equation}
The average $\left<\ldots\right>$ over the quenched pattern set can be
performed by the replica trick with the following replica symmetric
result:

\begin{equation}
\label{mthmoment:final2}
\left<y^{m}\right> = \int {\cal D}z\, \left[ \tanh\left( 
z\sqrt{\hat{R}_G}+\hat{R}_G \right) \right]^{m}\; ,
\end{equation}
where $\hat{R}_G$, which is determined by the saddle point equations
of Gibbs learning, cf.  eq.~\ref{spRG}, is found to be ${\hat R}_G=
{\cal F}^2(\sqrt{R_G})$.  Recognizing eq.~\ref{mthmoment:final2} as a
transformation of variables $y =
\tanh\left(z\sqrt{\hat{R}_G}+\hat{R}_G\right)$, with $z$ normally
distributed, one concludes \cite{comment} :
\begin{equation}
\label{px}
P(x)  =  \frac{\sqrt{R_G}}{\sqrt{2\pi\hat{R}_{G}}(1-R_G\, x^{2})}
\,\exp\left\{\frac{-1}{2\hat{R}_{G}}
\left[
\frac{1}{2}\ln\left( 
               \frac{1+\sqrt{R_G}\, x}{1-\sqrt{R_G}\, x}
               \right)
- \hat{R}_{G}
\right]^{2}\right\}\; .
\end{equation}

\begin{figure}[tbh]
\begin{center} 
\includegraphics{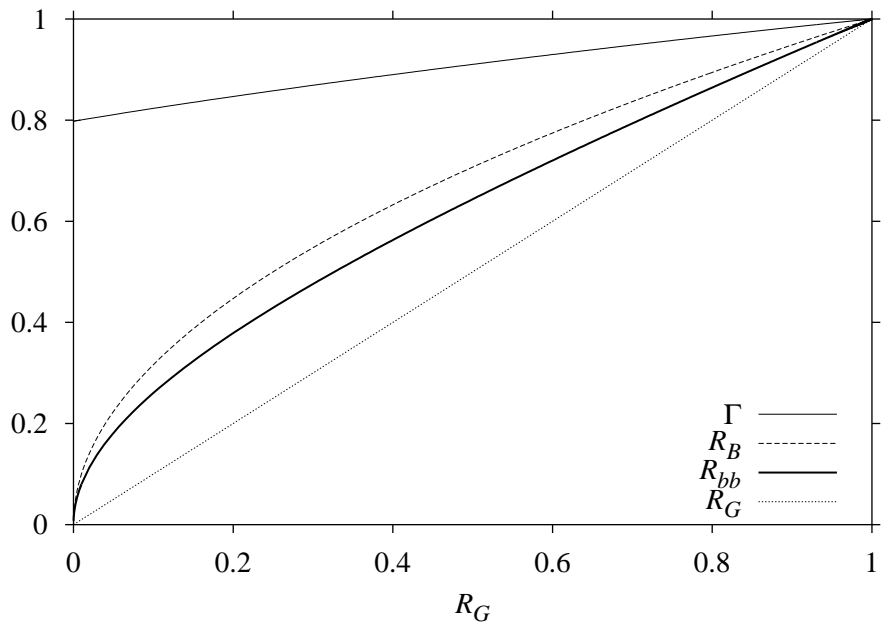}
\caption[short]{$\Gamma$, $R_{B}$ and $R_{bb}$ parametrized by
$R_{G}$, according to eqs.~\ref{spRG} and~\ref{Rbb}.  }
\label{Fig:overlaps}
\end{center}
\end{figure}

By applying eq.~\ref{rtilde}, for $g(x) = \mbox{sign}(x)$, with $P(x)$ 
given by eq.~\ref{px}, one finally obtains the following overlap 
$R_{bb}\equiv {\bf J}_{bb}\cdot{\bf B}/N$ of the best binary vector :

\begin{equation}
\label{Rbb}
R_{bb} = 1-2H\left(F_{B}^{-1}(R_{B})\right) = 1-2H\left({\cal F}\left( 
R_{B}\right)\right) \; ,
\end{equation}
where $H(x)\equiv \int_{x}^{\infty}Dt$.  Eq.~\ref{Rbb} is a central
result of this paper, providing an upper bound for the performance of
any binary vector.  The asymptotics of $R_{bb}$ can be obtained from
those of $R_{G}=R_B^2$, yielding

\begin{equation}
\label{Rbbasysmall}
R_{bb}\stackrel{R_{G}\to 0}{\simeq}  
\sqrt{ 
\frac{2R_{G}}{\pi}}
\end{equation}
in the poor performance regime, and an exponential behavior in the 
limit of $R_G \to 1$:

\begin{equation}
1-R_{bb}\simeq \frac{2}{\pi} (1-R_{G})
 \; .
\end{equation}
We note that another quantity of interest, the mutual overlap $\Gamma\equiv{\bf 
J}_{B}\cdot{\bf J}_{bb}/N$ between center of mass and best binary, can also be
evaluated quite easily, leading to the simple result
$\Gamma={R_{bb}}/{R_{B}}$.  In the limit $R_{G}\to 0$ one recovers 
$\Gamma\to\sqrt{2/\pi}$, which is the result for the 
overlap between a vector sampled at random from the $N$-sphere and its 
clipped counterpart.  $\Gamma$, $R_{B}$ and $R_{bb}$ are plotted as 
functions of $R_{G}$ in fig.~\ref{Fig:overlaps}.

We finally turn to the problem of a variationally optimized potential. 
In the case of a spherical prior, it was shown that the Bayes-optimal
performance can indeed be attained by a vector that minimizes this
potential~\cite{Kinouchi96a,VandenBroeck96,Buhot97b,Buhot98}.  We now
address the question of whether the same procedure is successful in
discrete space, a problem which has been also studied
in~\cite{Mattos97} for the supervised scenario.  Since ${\bf J}_{bb}$
is a unique optimal binary vector, one would like the desired
potential to satisfy both $R=R_{bb}$ and $q = 1$.  
Proceeding again from the free energy eq.~\ref{freeenergy} for a
general potential $V$, taking the limits $q\to 1$, $\beta\to\infty$
with finite $c \equiv \beta(1-q)$, and rescaling the conjugate
parameters ${\hat c}\equiv \hat{q}/\beta^{2}$, ${\hat y}\equiv
\hat{R}/\beta$, one obtains the following saddle point equations:

\begin{eqnarray} 
\label{spqto1}
R  =  1 - 2 H\left(\frac{\hat y}{\sqrt{\hat c}}\right)  &\;\;\;&
c  =  \sqrt{\frac{2}{\pi\hat c}}\exp\left(-\frac{{\hat y}^{2}}
{2{\hat c}}\right)  \\
{\hat c}  =  \frac{\alpha}{c^{2}} \int {\cal D}t\, X(t;R) 
\left[\lambda_{0}(t,c) - t\right]^{2} &\;\;\;&
\hat y  =  \frac{\alpha}{c}\int {\cal D}t\, 
Y(t;R)\left[\lambda_{0}(t,c) - t\right] \; ,\nonumber
\end{eqnarray}
where $\lambda_{0}(t,c) \equiv \Argmin_{\lambda}\left[ V(\lambda) + 
(\lambda -t)^{2}/2c \right]$.  The variational optimization of $R$ 
with respect to the choice of $V$ can now be performed as in 
refs.~\cite{Kinouchi96a,VandenBroeck96,Buhot97b,Buhot98} invoking the 
Schwarz inequality. 
We only quote the final result for the 
resulting overlap $R_{opt}$ at the minimum of this optimal potential: 

\begin{equation}
\label{Ropt}
R_{opt} = 1-2H\left({\cal F}(R_{opt})\right)\; .
\end{equation}
The important issue to be examined is whether or not $R_{opt}(\alpha)$
saturates the bound given by the best binary.  By comparison of
eq.~\ref{Ropt} with eq.~\ref{Rbb}, one immediately concludes that this
is {\em not possible}, as long as ${\cal F}$ is not a constant nor
singular, since $R_{opt} = R_{bb}$ would imply that $ {\cal F}(R_{bb})
= {\cal F}(R_{B})$, and $R_{bb}=R_{B}$ is excluded by the first
equality in eq.\ref{Rbb}.  In general one thus has that $R_{opt}\leq
R_{bb}$, since $\partial{\cal F}/\partial R \geq 0$.  The equality is
reached in asymptotic limits and for a special case (see below).  For
$R_{opt}\sim 0$ one has:

\begin{eqnarray}
\label{optV:smallalpha1}
\left<t\right>_{*}\neq 0 & \Rightarrow & R_{opt} \simeq 
\left|\left<t\right>_{*}\right|\sqrt{\frac{2\alpha}{\pi}} \\
\label{optV:smallalpha2}
\left<t\right>_{*} = 0 & \Rightarrow & 
R \left\{
\begin{array}{ll}
= 0, & \alpha\leq\alpha_{c} \\
\simeq \sqrt{C^{\prime} (\alpha-\alpha_{c})}, & \alpha\geq\alpha_{c}
\end{array}
\right.\; ,
\end{eqnarray}
where the critical value now is $\alpha_{c}\equiv \pi\alpha_{G}/2$.  
Furthermore, the approach $R_{opt}\rightarrow 1$ is identical to that 
of $R_{bb}$, $1-R_{opt} \simeq 1-R_{bb}$.  Therefore $V_{opt}$ is 
successful only in the asymptotic limits $\alpha\to 0$ and 
$\alpha\to\infty$.  Note that the second order phase transition in 
eq.~\ref{optV:smallalpha2} occurs at a larger value of $\alpha$ than 
for Gibbs learning.

The case ${\cal F}(R)$ independent of $R$, implying $R_{opt} =
R_{bb},\; \forall \alpha$, arises in a simple Gaussian scenario with a
linear function $U$~\cite{Copelli99c}.  In this case the best binary
corresponds to clipped Hebbian learning.  This seems to be the only
case in which minimization of an optimal potential reproduces the best
binary vector.  We conclude that an optimal potential saturating the
$R_{bb}$ bound with $q\to 1$ cannot be constructed, in general.  It
motivates the search for alternative methods in discrete optimization. 
The main issue is to find new ways to incorporate information about
the binary nature of the symmetry breaking vector, other then simply
imposing the same binary constraint in the solution space.  An
interesting approach would be to try to construct a suitable potential
for the continuous center of mass ${\bf J}_{B}$ from which the best
binary could be obtained by clipping.  Whether such an approach is
possible will be answered in future work.

\end{sloppypar}

The authors would like to thank Nestor Caticha for useful discussions
and the organizers of the International Seminar on ``Statistical
Physics of Neural Networks'' (1999), held at the Max-Planck Institut
f\"ur Physik komplexer Systeme (Dresden), during which part of this
work was accomplished.  We also acknowledge support from the FWO
Vlaanderen and the Belgian IUAP program (Prime Minister's Office).

%
%
%


\begin{thebibliography}{10}
\bibitem{comment}
Eq.~\ref{px} is consistent with the fact that the overlap for the
center of mass cannot be improved by a transformation of its
components.  Indeed the transformed overlap $\tilde{R}$ in
eq.~\ref{rtilde} is maximized~\cite{Schietse95} by setting $g_{opt}(x)
= (P(x)-P(-x))/(P(x)+P(-x))$, reducing for (\ref{px}) to
$g_{opt}(x)\sim x$.
\end{thebibliography}
\end{document}